\documentclass[journal]{IEEEtran}
\ifCLASSINFOpdf
\usepackage[pdftex]{graphicx}
\else
\fi
\usepackage{amsmath}
\usepackage{array}

\ifCLASSOPTIONcompsoc
 \usepackage[caption=false,font=normalsize,labelfont=sf,textfont=sf]{subfig}
\else
 \usepackage[caption=false,font=footnotesize]{subfig}
\fi
\begin{document}
%
\title{Fabrication of the impedance-matched Josephson parametric amplifier and the study of the gain profile}
%
%
%
\author{Rui~Yang and Hui~Deng
\thanks{Manuscript created September 2019. revised November 2019. This work
was supported by the University of Science and Technology of China.}
\thanks{Rui Yang is with Hefei National Laboratory for Physical Sciences at Microscale and Department of Modern Physics, University of Science and Technology of China, Hefei, Anhui 230026, China,
and also with the Shanghai Branch, CAS Center for Excellence and Synergetic Innovation Center in Quantum Information and Quantum Physics, University of Science and Technology of China, Shanghai 201315, China, e-mail:smart8@ustc.edu.cn }
\thanks{Hui Deng is with Hefei National Laboratory for Physical Sciences at Microscale and Department of Modern Physics, University of Science and Technology of China, Hefei, Anhui 230026, China,  e-mail:huid@ustc.edu.cn }

}

%
%

\markboth{TAS-2019-0211}%
{Yang \MakeLowercase{\textit{et al.}}: Fabrication of the impedance-matched Josephson parametric amplifier and the study of the gain profile}
%



\maketitle



\begin{abstract}
We designed and fabricated an impedance-matched Josephson parametric amplifier (JPA) working in the flux-pump mode for the broadband amplification of microwave signals. We developed a simple fabrication method suitable for a small cleanroom. We studied the phase response, as well as the gain, as a function of signal frequency and pump power at various pump frequencies. The phase response can be explained with the behavior of the non-linear Duffing oscillator. The observed decrease of the resonance frequency as the pump power increases, as well as the emergence of an unstable bifurcation zone, are the characteristic non-linear behavior of the Duffing oscillator. The features of the gain profile in the stable zone can be explained with a model adapted from the theoretical model for the two-dimensional gain profile of an impedance-matched, current-pumped JPA. Over 17 dB gain with a bandwidth more than 300 MHz (centred around 6.45 GHz) was observed. Our impedance-matched JPA is used for improving the fast readout fidelity of transmon qubits.
\end{abstract}


%
\IEEEpeerreviewmaketitle

\section{Introduction}
%
%
%
%
\IEEEPARstart{Q}{uantum} computers hold the promise for solving many problems that can not be solved efficiently with conventional computers\cite{Feynman,shor,Grover,Grover2}. It can provide an efficient way to simulate quantum mechanical systems\cite{Feynman}. In some applications, quantum algorithms also show great advantage over conventional algorithms\cite{shor, Grover,Grover2}. The proof-of-concept demonstration of quantum computation and quantum simulation on small scale quantum computers has been realized\cite{surfacecode2, expshor, expprimefactor, expsearch, expsearch2,scclassifier,SchusterBoseinsulator}. The promising prospect of building a large quantum computer of practical use draws the attention of many institutions. Many countries join the race towards a practical quantum computer. Within the last several years, progress in the study of superconducting quantum computers is impressive. For example, Google built a 72-qubit quantum computer; IBM built quantum processors with up to 50 qubits and put several quantum processors on the cloud. Intel also built quantum processors with more than 50 qubits. A key issue in the development of a practical multi-qubit quantum computer is the high-fidelity readout of the signals from multiple qubits. A high distinguishability between the qubit states is crucial for various operations on the quantum computer, such as running quantum algorithm, error correction and quantum feedback experiments\cite{ro1,ro2,ro3,ro4,ro5}. The microwave signals carrying information of the qubits are very weak. Amplification is thus required. The commercial microwave HEMT amplifier adds lots of noise during amplification, the equivalent noise temperature is several Kelvins\cite{ro6,ro7sank}, thus degrading the distinguishability between quantum states. To improve the readout fidelity, an amplifier with the lowest possible added noise is favourable. Josephson parametric amplifier (JPA) is the most popular system that can provide such a low noise level\cite{ro7sank,Mutusnarrow,Caves}. To read many qubits simultaneously, a large bandwidth for the Josephson parametric amplifier is also required. The simplest JPA is a non-linear resonator composed of a superconducting quantum interference device (SQUID) loop shunted by a big capacitor. However, the bandwidth of such a resonant structure is quite small, thus making it a narrow-band JPA. A way to increase the bandwidth of this type of JPA is to incorporate a gradual impedance transformer into the microwave feedline, thus lowering the quality factor of the resonant structure without introducing too much reflection\cite{Mutus}. The gain profile of such an impedance-matched JPA differs from that of a narrow-band JPA. The characterization and understanding of this gain profile can help in choosing the best working zone for the qubit readout task.



In this work, we first designed and fabricated an impedance-matched Josephson parametric amplifier working in the flux-pump mode\cite{2008APL}. The fabrication process we developed simplifies the fabrication of the impedance transformer part. The process is suitable for a small cleanroom without etching facilities for the dielectrics.  We studied the phase and amplitude of the microwave scattering parameter, as well as the gain, as a function of signal frequency and pump power, at various pump frequencies. The phase response reflects the behavior of the non-linear Duffing oscillator. The observed decrease of the resonance frequency as a function of the pump power, as well as the emergence of an unstable bifurcation zone (which also has an impact on the amplitude response), are characteristic for a non-linear Duffing oscillator. The observed amplitude or gain response has two zones. When the pump power is below some critical level, the system is in the stable zone. When the pump power is larger than a critical value, the system entering the bifurcation zone, the amplitude or gain response is unstable and a sudden change usually occurs. We adapted a theoretical model\cite{TanaRoy} for the two-dimensional (2d) gain profile of an impedance-matched current-pumped JPA to explain the observed 2d gain profile in the stable zone. The theoretical model\cite{TanaRoy} was originally used for describing the gain profile of JPAs working in the current-pump mode, which indicates that the environmental impedance can influence the details of the gain profile of a JPA. With an appropriate environmental impedance, the theoretical model captures the features of our data, such as the emergence of a gain hot zone with two branches around the resonance frequency of the JPA.  Based on the gain profile, we propose that the best working zone is the merging point of the gain hot zone before the emergence of the bifurcation zone, which gives a large bandwidth and good gain. Over 17 dB gain with a bandwidth larger than 300 MHz centred around 6.45 GHz was observed. The impedance matched JPA is used in our experiments for improving the fast readout fidelity of the transmon qubits.
\section{Design and Device Fabrication}
\begin{figure}[!htb]
\subfloat[\label{fig8p9(a)}]

  {\includegraphics[width=1.0\linewidth,height=6cm]{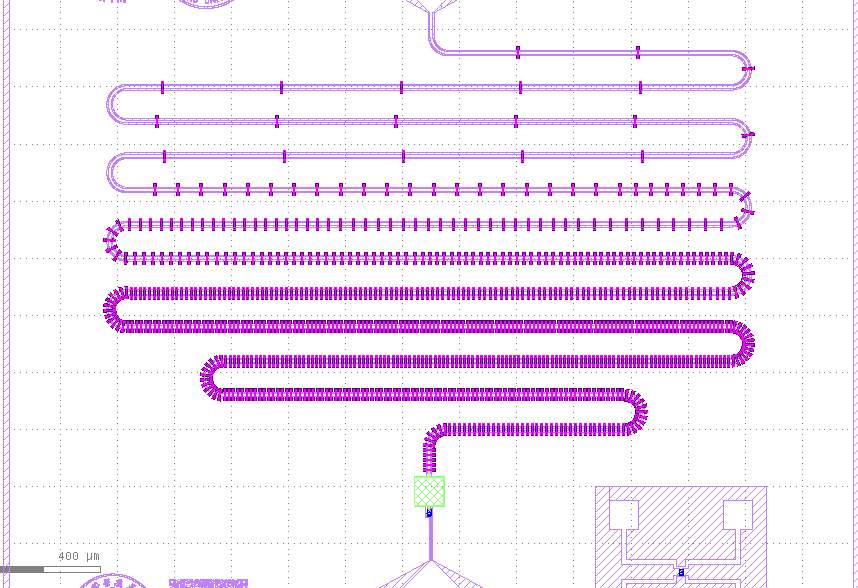}}\hfill
\subfloat[\label{fig8p9(b)}]

  {\includegraphics[width=1.0\linewidth,height=6cm]
  {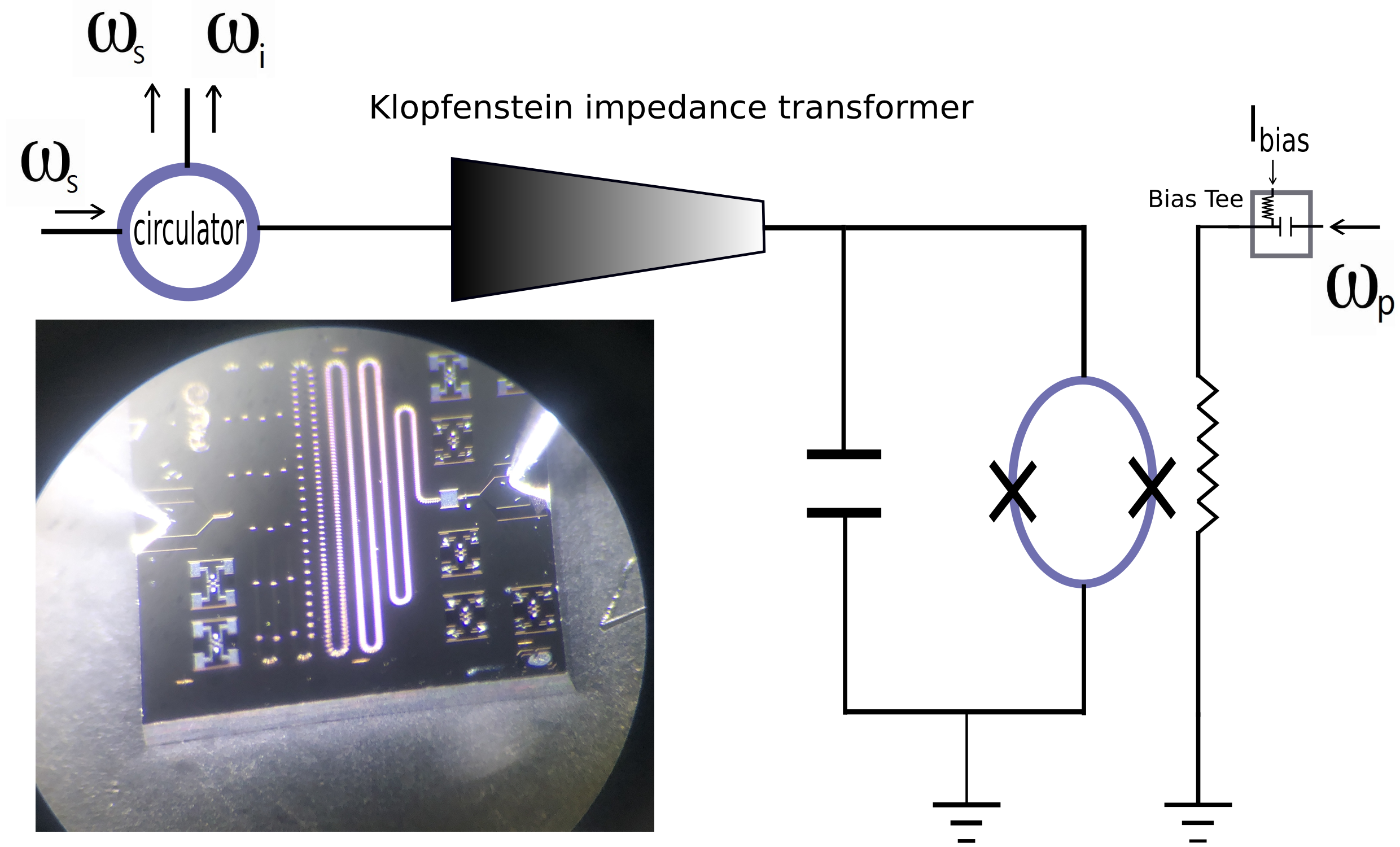}}

\caption{(a) The design of the impedance transformer (the meandering line represents the CPW transmission line, the magenta structures represent the small crossovers). The density of the small crossovers modified the impedance along the line. The denser the crossovers, the smaller the impedance of the segment of transmission line will be. (b) The circuit schematic of the JPA (the inset is a photo of our JPA). The circle with two crosses represents the SQUID loop with two Josephson junctions, shunted by a capacitor. The line connected with 'Bias Tee' represents the flux bias line, which tunes the flux penetrating the SQUID loop. In addition to the DC bias $I_{bias}$, a RF bias serving as a pump tone is also applied on this line with the help of a Bias Tee. $\omega_{p}$ is the frequency of the pump tone. $\omega_{s}$ is the frequency of the signal to be amplified. $\omega_{i}$ is the frequency of the idle tone generated during the amplification. The three frequencies satisfy $\omega_{p}=\omega_{s}+\omega_{i}$. In the operation mode, $\omega_{s}=\omega_{i}$, thus $\omega_{p}=2\omega_{s}$. }
\label{fig8p5}
\end{figure}

The impedance-matched JPA was designed and fabricated at the Nanofabrication Facilities at Institute of Physics in Beijing, University of Science and Technology of China in Hefei, and National Center for Nanoscience and Technology in Beijing. The impedance transformer is a compact Klopfenstein filter which gradually changes the impedance of the transmission line from 50 $\Omega$ to a smaller value (about 15 $\Omega$, limited mainly by the dielectric constant of the crossover capacitors and the size of the JPA), thus increasing the bandwidth as well as the saturation power of the JPA to the desired values, since both the bandwidth and the saturation power are inversely proportional to the impedance seen by the nonlinear resonator composed of the Josephson junction and the shunt capacitor\cite{Mutus}. The large shunt capacitor is designed to be around 4 pF, the inductance of the Josephson junction is designed to be around 70 pH, the corresponding room temperature junction resistance is around 62 $\Omega$. The design of the impedance transformer and the circuit illustration can be found in Fig.~\ref{fig8p5}.



 The device fabrication involves many processes. In the first step, a thin Al film 100 nm thick was deposited on a cleaned silicon wafer (with a 1 micron thick SiO$_{2}$ layer). In the second step, a layer of photoresist S1805 was spin-coated and baked. Then, lithography was performed by a laser writer to form the etch mask. After that, wet etching was performed to define the co-planar wave-guide (CPW) by removing some Al from the designed area. In the third step, photoresist LOR5A and S1805 were spin-coated and lithography by a laser writer was performed again to define the regions for the dielectrics. Then, a layer of 200 nm thick CaF$_{2}$ layer was deposited and liftoff was performed to form the dielectric layer. 
 In the fourth step, LOR5A and S1805 were spin-coated again and lithography by a laser writer was performed again to define the regions for the top electrodes of the shunt capacitor and the crossovers on the impedance transformer.
 Then, a layer of Al was deposited again to form the top electrodes. There is no etching for the dielectrics in our fabrication steps. This simplified fabrication process saves the steps of etching dielectrics. The Josephson junction region is defined by the electron beam lithography on a MAA/PMMA ebeam resist mask. The Josephson junctions were deposited in a deposition machine with the double angle evaporation method\cite{Dolanbridge}.
\section{Modulation curves}
\begin{figure}[!htb]
\centering
  {\includegraphics[width=1.0\linewidth, height=6.5cm]{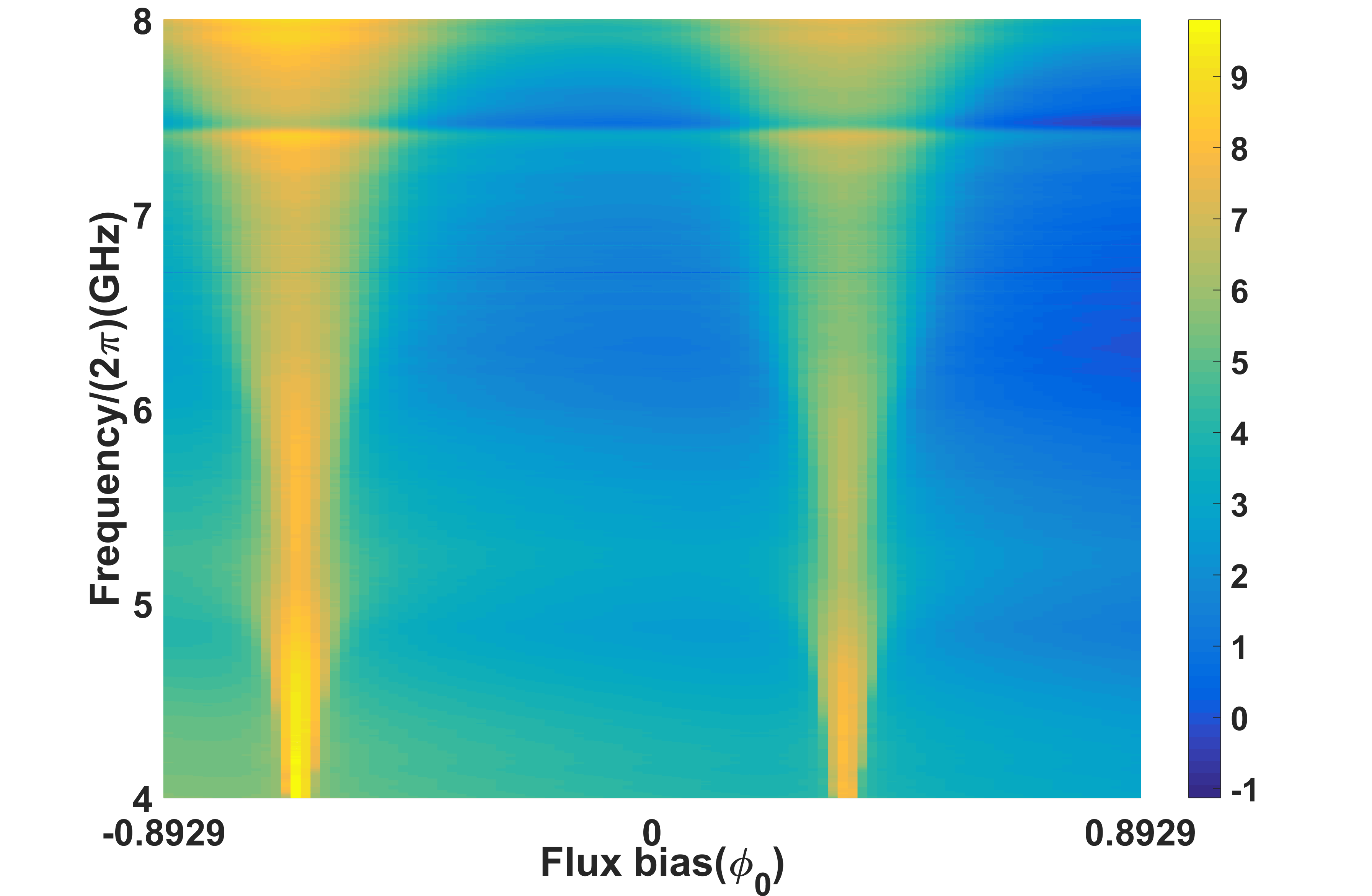}}\hfill 
\caption{Flux modulation of the phase of the reflected microwave (phase of the reflected microwave as a function of the signal frequency and the flux bias, measured by a vector network analyser). The boundary between the bright and dark regions indicates the resonance frequencies.}
\label{fig8p9}
\end{figure}

The phase angle of the microwave reflected from the JPA can be modulated by the flux penetrating the SQUID loop. Fig.~\ref{fig8p9} shows the phase of the reflected microwave as a function of the microwave frequency and flux bias. The observed phase modulation can be modelled by $angle=\arctan (\frac{2Z_{0}}{\omega L_{j}(\phi, I_{c})} \frac{((1-\omega^{2}L_{j}(\phi, I_{c})C)^2)^2-Z_{0}^{2} }{1-\omega^{2}L_{j}(\phi, I_{c})C })$, with $L_{j}(\phi,I_{c})=\frac{\phi_{0}}{2\pi I_{c} |\cos(\pi \phi/\phi_{0})| }$, where $\omega$ is the frequency of the probe tone, $\phi_{0}$ is the flux quantum (about $2.07\times10^{-15}$ Wb), $Z_{0} \approx 15$ $\Omega$ is the transformed characteristic impedance of the transmission line, $I_{c}$ is the zero-field critical current, $\phi$ is the flux penetrating the SQUID loop, $C$ is the shunt capacitance\cite{phasemodulation}. The zero-field critical current $I_{c}$ and shunt capacitance can be obtained by fitting theory to the observed phase modulation curves of the JPA. The extracted $I_{c}$ is about 4.5 $\mu$A. The extracted $C$ is about 3.5 pF. These values are close to the designed values.


\section{Two dimensional Gain profile}
\begin{figure}[!htb]
\centering
  {\includegraphics[width=1.0\linewidth, height=4cm]{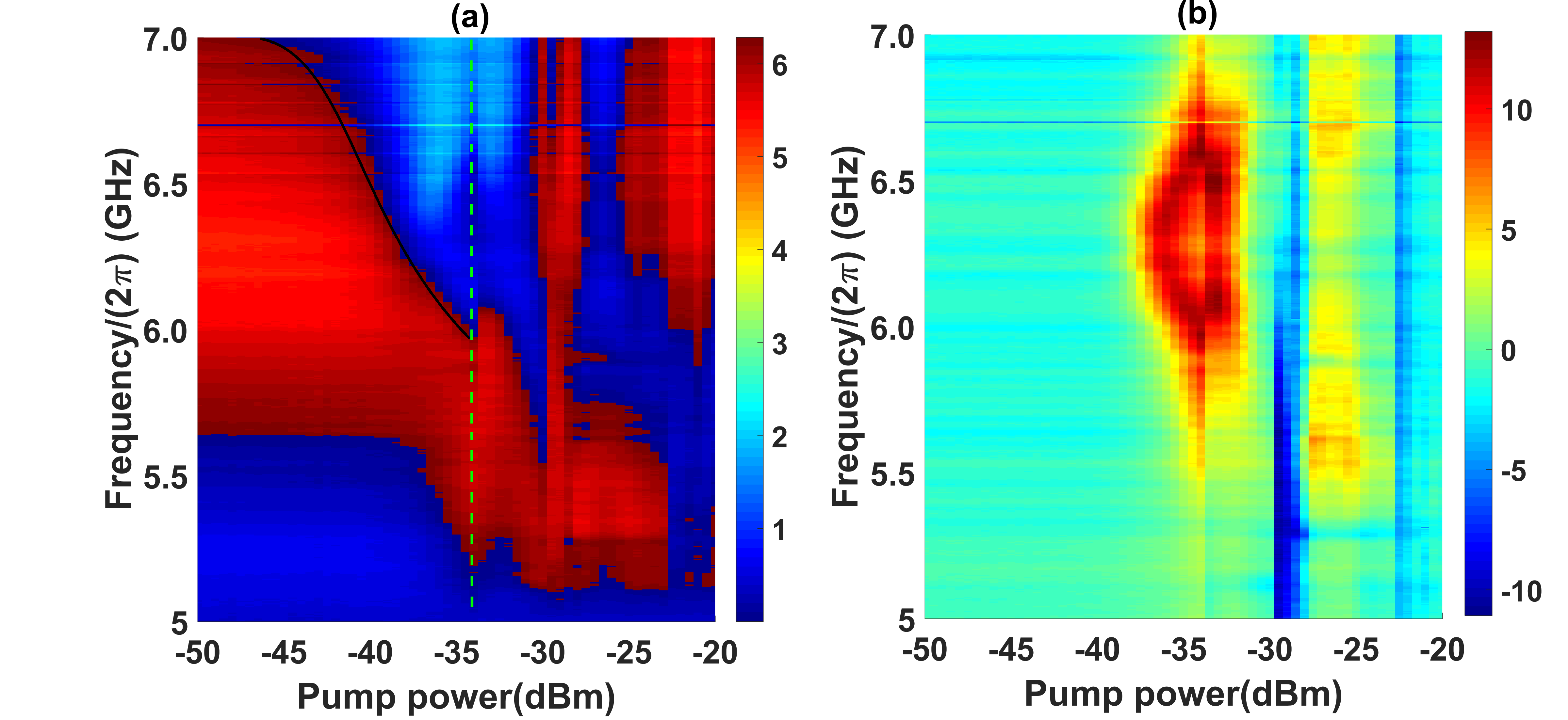}}\hfill 

\caption{(a) An example of the phase response (phase as a function of signal frequency and pump power at the RF input to the JPA) for pump frequency around 12.6 GHz and flux bias around 0.3$\phi_0$. The black curve indicates the decrease of the resonance frequency as a function of pump power. The green dashed line indicates the beginning of the bifurcation zone. (b) The corresponding amplitude response of (a), the entering of the bifurcation zone also manifests itself here, the abrupt change when pump power is around -34 dBm indicates the beginning of the bifurcation zone.}
\label{fig12p3}
\end{figure}

After studying the phase modulation of the JPA devices caused by the external flux bias, we turned on the flux pump and characterized the 2d gain profile (gain as a function of the signal frequency $\omega_s$ and pump power at the sample, at various pump frequencies and flux biases) in detail. An example of the measured 2d gain profile can be found in Fig.~\ref{fig12p3}(b).

This 2d gain profile shows a phase diagram of the JPA, if we look at the phase information of the microwave. Take Fig.~\ref{fig12p3}(a) as an example, at a low pump power,
the boundary indicated by the black curve represents the pump-power dependence of the resonance frequency of the JPA. We can see the resonance frequency shifts to lower values as the pump power increases. This phenomenon can be explained by the non-linearity of the Duffing oscillator\cite{Vijaythesis,Casteno}.  Above a threshold pump power, the phase changes suddenly. This phenomenon could also be explained by the physics of the non-linear Duffing oscillator. Above some critical value of the pump power, the Duffing oscillator enters a bifurcation zone. The resonance amplitude will become multi-valued in this zone\cite{Vijaythesis,Casteno}, leading to the instability of the system. The bifurcation zone also manifests itself as a sudden change of the gain in the corresponding amplitude or gain plot (Fig.~\ref{fig12p3}(b)). Comparing the amplitude or gain plot (Fig.~\ref{fig12p3}(b)) with the phase plot (Fig.~\ref{fig12p3}(a)), we can see some gain hot zones in the different regions defined by the phase response. Further experiments show the gain hot zone lies in the region between the black curve and the bifurcation zone is stable (useful for readout purpose of qubit states). The stable gain hot zone usually occurs in the region not far from the onset of the bifurcation zone. This is because the cavity photon number changes abruptly at the bifurcation point, thus having the maximum parametric conversion effect\cite{Vijaythesis,Casteno}. Therefore, the region between the black curve and the bifurcation zone is where we should focus to search for the optimal stable gain hot zone. 

Let us focus on the stable gain in the stable region. We can see the gain hot zone generally tracks the curve representing the pump-power dependence of the resonance frequency. Thus the gain hot zones for a higher pump frequency are located at a smaller pump power, because the resonance frequency increases at a lower pump power. The gain hot zone usually has two branches. Sometimes they merge into a broad plateau. A theoretical model\cite{TanaRoy} for the 2d gain profile of an impedance-matched current-pumped JPA can be adapted to describe the observed 2d gain profile of our JPA. The gain profile of an impedance-matched current-pumped JPA can be calculated based on the differential equation of the circuit system or the Hamiltonian of the circuit system. The differential equation for the current-pump mode reads\cite{TanaRoy}:



\[\frac{d^{2} \delta_{s}(t) }{dt^{2}} + \kappa_{0} \frac{d \delta_{s}(t) }{dt} + \Omega_{p}^{2}(1-\epsilon)(1-\frac{\epsilon}{1-\epsilon} \sin(2\Omega_{d} t -2\theta) )\delta_{s}(t)= A_{s}(t)\] where $\delta_s$ is the phase parameter on the Josephson junction, $\Omega_p$ is the resonant frequency, $\Omega_d$ is the drive frequency in the current-pump mode, $A_s$ is the signal amplitude, $\epsilon$ is controlled by the drive amplitude. More details can be found in reference\cite{TanaRoy}.



Comparing it to the differential equation of the flux-pump mode:



\[\frac{d^{2} \delta_{s}(t) }{dt^{2}} + \kappa_{0} \frac{d \delta_{s}(t) }{dt} + \Omega_b^{2} (1-\eta\cos(\Omega_{pump} t) )\delta_{s}(t) = A_{s}(t)\] where $\Omega_{b}^{2}=\frac{2I_{c}\times 2\pi}{C\phi_{0}}\cos(\frac{\pi}{\phi_{0}}\Phi_{b} )$,$\eta=\tan(\frac{\pi}{\phi_{0}}\Phi_{b})\frac{\pi}{\phi_{0}}\Delta\Phi$, $\Phi_b$ is the flux bias, $\Delta\Phi$ is the amplitude of the flux pump, $C$ is the shunt capacitance and $I_c$ is the critical current. We can see these two equations are basically identical, with some replacements: $\Omega_{pump} = 2\Omega_{d}$,     $\Omega_{p}^{2}(1-\epsilon)=\frac{2I_{c}\times 2\pi}{C\phi_{0}} \cos(\frac{\pi}{\phi_{0}}\Phi_{b})$ and $\epsilon=\frac{1}{1+\frac{1}{\tan(\frac{\pi}{\phi_{0}}\Phi_{b})\frac{\pi}{\phi_{0}}\Delta \Phi }}$.






Therefore, the differential equation for the flux-pump mode is similar to that of the current-pump mode. We just need to do some substitution, replacing some terms in the parametric oscillating term in the current-pump mode with their counterparts in the flux-pump mode. In other words, there is some one-to-one correspondence between the terms of the differential equation in the current-pump mode and the terms of the differential equation in the flux-pump mode. Therefore, we can directly apply the gain formula of the current-pump mode to the flux-pump mode by simply converting the pump strength in the current-pump mode to the flux pump power in the flux-pump mode.  

The resulted gain is $g=|1-\kappa_{1}\chi_{11}|^{2}$. $\kappa_{1}$ is related to the real part of the environmental admittance and $\chi_{11}$ is an element of the susceptibility matrix. Details can be found in reference \cite{TanaRoy}. 

\begin{figure}[!htb]
\centering
  {\includegraphics[width=1\linewidth, height=6.5cm]{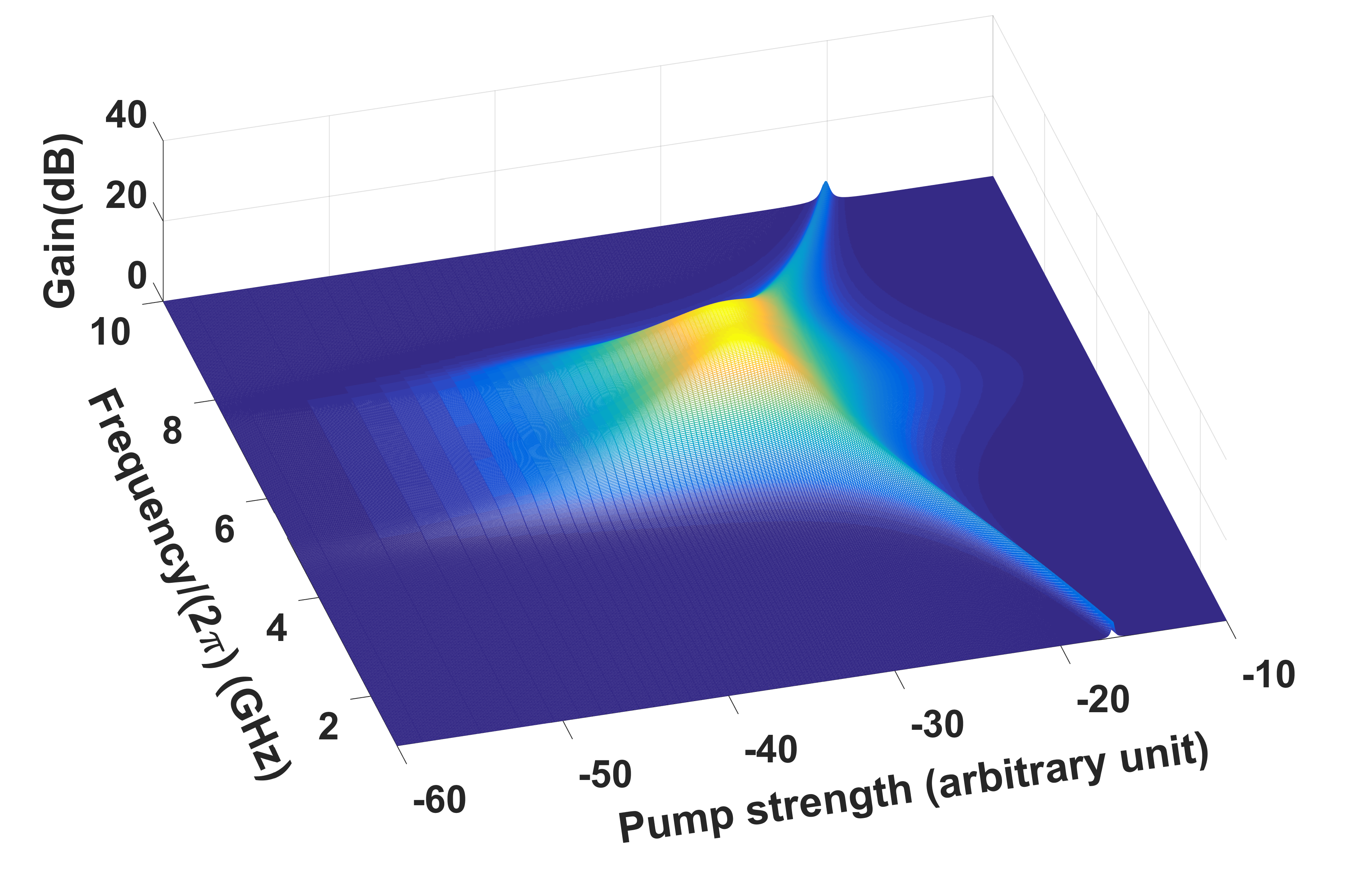}}\hfill 

\caption{Calculated 2d gain profile for the pump frequency around 12.6 GHz and flux bias at 0.3$\phi_0$. The calculated gain as a function of pump strength and signal frequency qualitatively matches with the observed gain profile in the stable zone. Since the theory works in the stable zone, this calculation doesn't capture the abrupt change in the bifurcation zone.}
\label{fig8p2}
\end{figure}

With the critical current and shunt capacitance extracted from the phase modulation curves and an appropriate environmental impedance as the input, the theoretical model gives a 2d gain profile which qualitatively matches with the data. The calculated gain profile can be found in Fig.~\ref{fig8p2}. The model captures the features of the data, such as the emergence of a gain hot zone with two branches around the resonance frequency of the JPA\cite{TanaRoy}. The branching behavior in the gain hot zone depends on the environmental impedance. For a non-optimal environmental impedance, the gain profile has two branches, related to the normal modes of the JPA and the environment\cite{TanaRoy}. For the optimal environmental impedance, the two branches merge, giving rise to a broad plateau. The gain hot zone usually gets disturbed by the emergence of the bifurcation zone and having a distorted shape extended into the bifurcation zone.

\begin{figure}[ptb]
\centering
\includegraphics[width=1.0\linewidth]{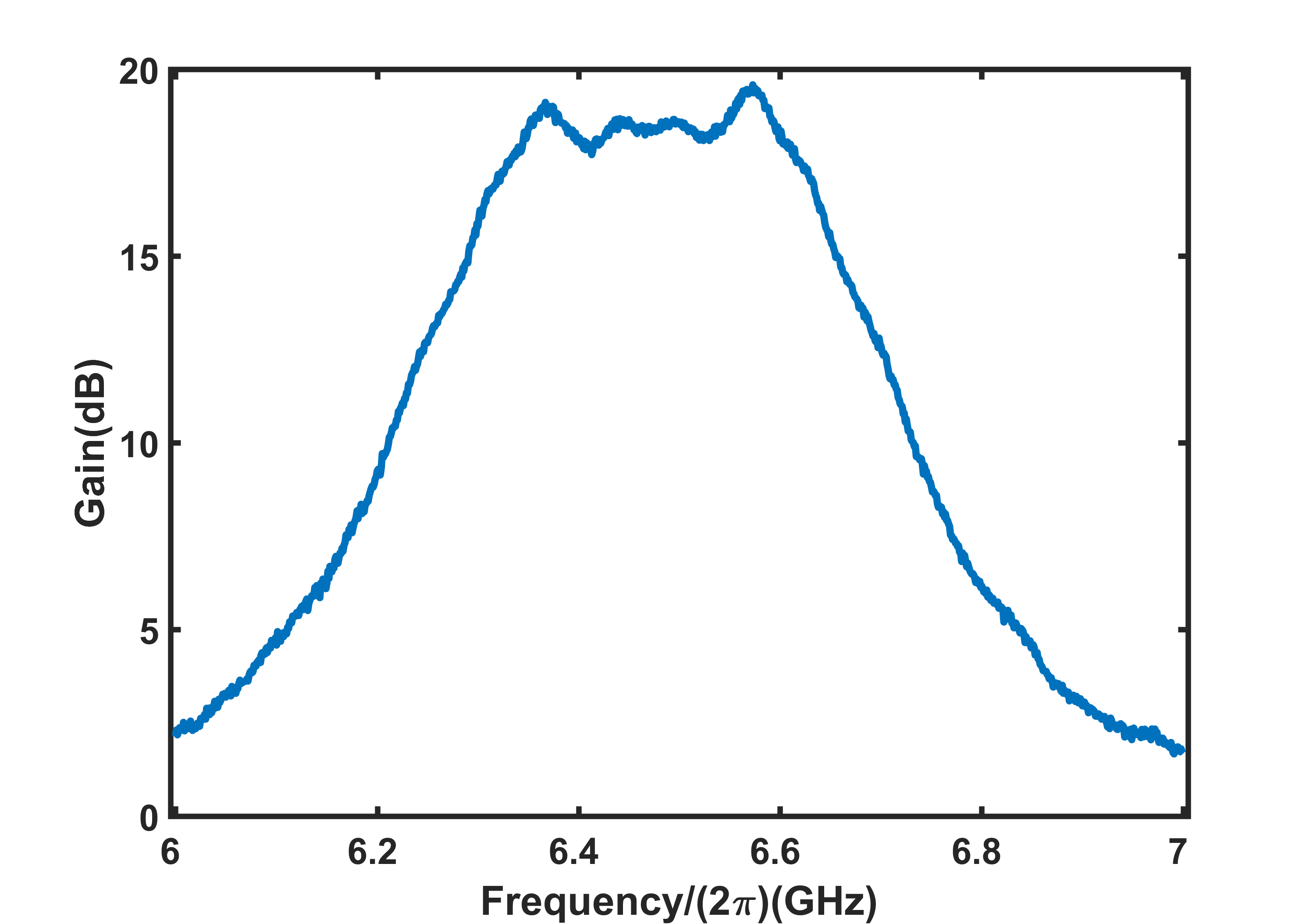}
\caption{Bandwidth of the JPA. This figure shows the gain as a function of the signal frequency. We can see that gain over 17 dB can be reached in a frequency range of around 300 MHz. The pump frequency is around 12.9 GHz. The flux bias is around 0.3$\phi_0$. The pump power is around -40 dBm at the RF input to the JPA. }
\label{fig8p12}
\end{figure}

Based on the 2d gain profile, we propose that the best working zone is near the merging point of the two branches of the gain hot zone before the emergence of the bifurcation zone, which gives a large bandwidth and good gain. In our JPA, larger than 17 dB gain with a bandwidth larger than 300 MHz (centred around 6.45 GHz) was observed, see Fig.~\ref{fig8p12} (the pump frequency is around 12.9 GHz. The flux bias is around 0.3$\phi_0$. The pump power is around -40 dBm at the RF input to the JPA). The large bandwidth with a decent gain is ideal for the readout task of the multi-qubit quantum chips.

\section{Saturation power and noise temperature}
Besides the gain profile, we also characterized the saturation power and noise temperature of our JPA. The gain decreases at higher signal power. The saturation power can be characterized by the 1 dB compression point, the value of the signal power where the gain decreases by 1 dB. The 1 dB compression power in the middle of our best gain hot zones (Fig.~\ref{fig8p12}) is about -110 dBm. The noise temperature measures how much noise is added during the amplification process. In our setup, the noise temperature of our JPA in the region with gain larger than 15 dB is about 300 mK, which is close to the quantum limit. More details about the noise temperature of the type of JPA made by us can be found in the literature\cite{HuangKQ}.
\section{Qubit readout}
\begin{figure}[!htb]
\subfloat[\label{fig8p4(a)}]

  {\includegraphics[width=1.0\linewidth,height=6cm]{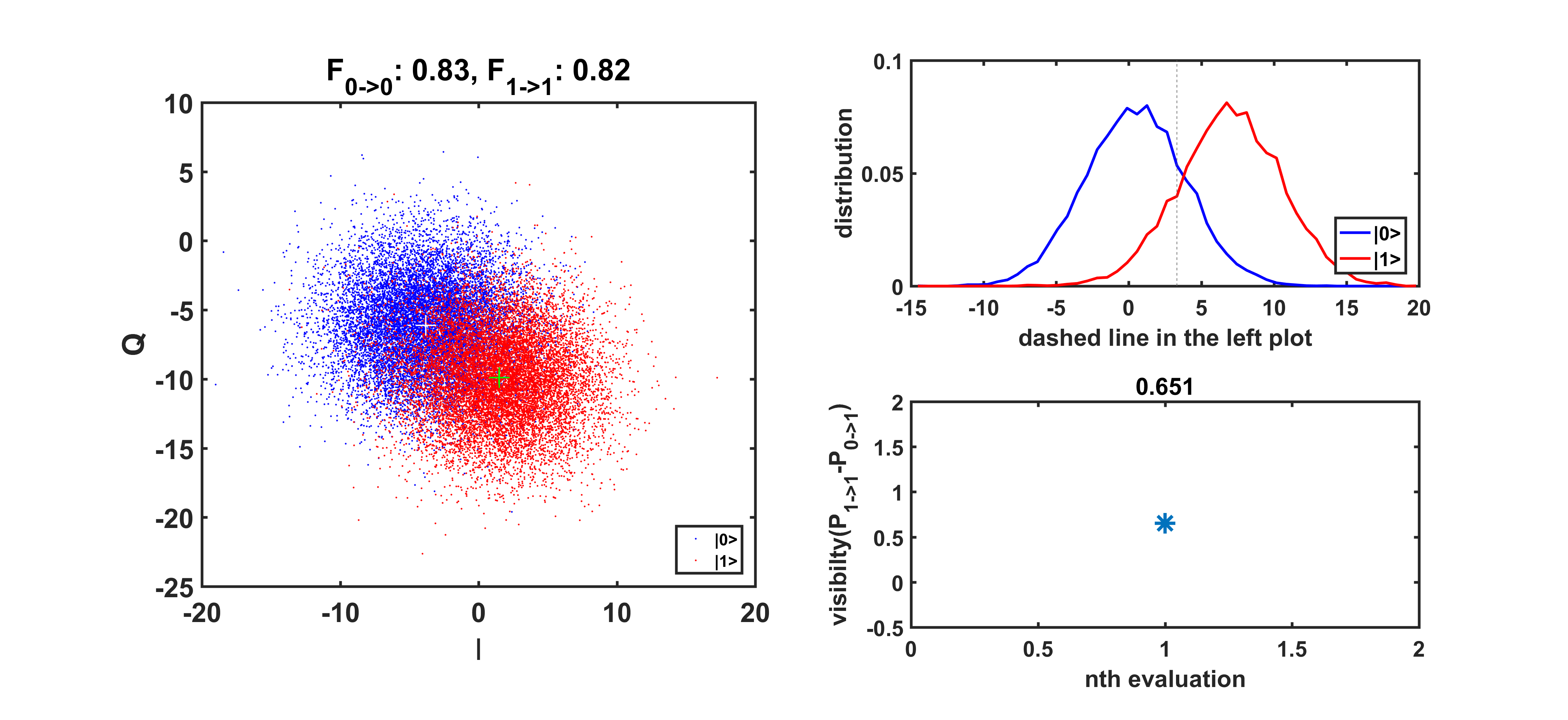}}\hfill
\subfloat[\label{fig8p4(b)}]

  {\includegraphics[width=1.0\linewidth,height=6cm]
  {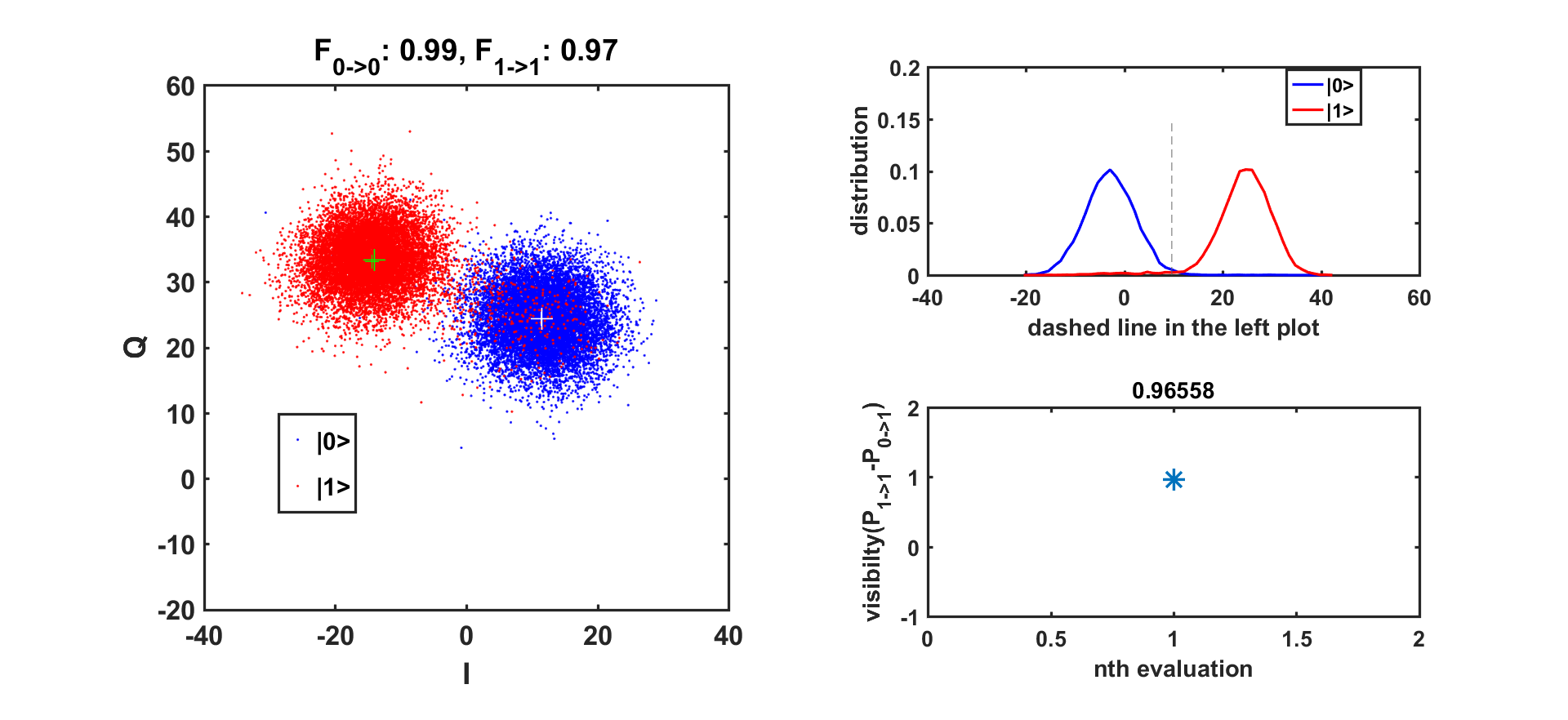}}


\caption{(a) Readout fidelity with JPA off. The fidelity for the ground state and excited state are about 0.83 and 0.82, respectively.  (b) Readout fidelity with JPA on. The fidelity for the ground state and excited state are now about 0.99 and 0.97, respectively.}
\label{fig8p4}
\end{figure}

Our impedance-matched JPA is used in our experiments for improving the fast readout fidelity of transmon qubits. Fig.~\ref{fig8p4} shows the improvement in the readout fidelity with the JPA. Fig.~\ref{fig8p4}(a) shows the fidelity without the JPA in action, Fig.~\ref{fig8p4}(b) shows the fidelity with the JPA turned on. We can see that our JPA improved the readout fidelity a lot. In the experiments on the superconducting qubit chips, up to 6 qubits can be read simultaneously with the JPA\cite{yys24}. The large bandwidth with a decent gain and a decent saturation power is ideal for the readout task of our multi-qubit experiments.

\section{Conclusions}
In summary,  we designed and fabricated an impedance-matched Josephson parametric amplifier working in the flux-pump mode. The simple fabrication process is suitable for a small cleanroom without the etching facilities for dielectrics. We studied the 2d gain profile in detail. We adapted a theoretical model for the 2d gain profile of an impedance-matched current-pumped JPA to describe the observed 2d gain profile of our JPA. With an appropriate environmental impedance, the theoretical model captures the features of the gain profile, such as the emergence of a gain hot zone with two branches around the resonance frequency of the JPA. Non-linear behaviour was also observed, such as the shift of the resonance frequency at a higher pump power (Duffing shift), as well as the emergence of the bifurcation zones (unstable regions with a sudden drop of the gain) at a higher pump power. Based on the gain profile, we propose that the best working zone is the branching point of the gain hot zone in the stable zone right before the emergence of the bifurcation zone, which gives a large bandwidth and good gain. Gain Over 17 dB with a bandwidth larger than 300 MHz (centred around 6.45 GHz) was observed. The impedance-matched JPA is used in our experiments for improving the fast readout fidelity of the transmon qubits.

\section*{Acknowledgment}

This work was partially carried out at the USTC Center for Micro and Nanoscale Research and Fabrication. This research was supported by the National Basic Research Program (973) of China (Grant No. 2017YFA0304300), the Chinese Academy of Sciences, Anhui Initiative in Quantum Information Technologies, Technology Committee of Shanghai Municipality, NSFC (Grants No. 11574380, No. 11774406, No. U1530401). National Key Research and Development Program of China (Grant No. 2016YFA0302104, No. 2016YFA0300600), Strategic Priority Research Program of Chinese Academy of Sciences (Grant No. XDB28000000), China Postdoctoral Science Foundation (Grant No. 2018M640055), and Beijing Science Foundation (Grant No. Y18G07). The authors also thank QuantumCTek Co., Ltd. for supporting the fabrication and the maintenance of room temperature electronics.

\ifCLASSOPTIONcaptionsoff
  \newpage
\fi



%
 \bibliographystyle{IEEEtran}
 \bibliography{IEEEabrv,IEEEexample}

\end{document}